\documentstyle{l-aa}
\begin{document}

  \thesaurus{  09.08.1;    % HII regions 
               11          % Galaxies
              (11.01.1;    % Galaxies: abundances
               11.09.4;    % Galaxies: ISM
               11.09.5;    % Galaxies: irregular
               11.19.2) }  % Galaxies: spiral

   \title{On the oxygen abundance determination in HII regions}

   \subtitle{The problem of the line intensities -- oxygen abundance calibration}

   \author{ L.S. Pilyugin }

  \offprints{L.S. Pilyugin }

   \institute{   Main Astronomical Observatory
                 of National Academy of Sciences of Ukraine,
                 Goloseevo, 03680 Kiev-127, Ukraine, \\
                 (pilyugin@mao.kiev.ua) }
                 
   \date{Received ; accepted }

\maketitle

\markboth {L.S.Pilyugin: on the oxygen abundance determination in HII regions}{}

\begin{abstract}

The problem of the line intensities -- oxygen abundance calibration has been
considered. We confirm the idea of McGaugh (1991) that the strong oxygen lines
($[OII] \lambda \lambda 3727, 3729$ and $[OIII] \lambda \lambda 4959, 5007$)
contain the necessary information for determination of accurate abundances in
low-metallicity (and may be also in high-metallicity) HII regions.
It has been found that the excitation parameters $p_{3}$ 
or $p_{2}$ (which are defined here as contributions of the radiation in 
$[OIII] \lambda \lambda 4959, 5007$ lines and in 
$[OII] \lambda \lambda 3727, 3729$ lines to the "total" oxygen radiation 
respectively) allow to take into account the variations in $R_{23}$ values
among HII regions with a given oxygen abundance.
Based on this fact a new way of the oxygen abundance determination in
HII regions (p -- method) has been constructed and corresponding 
relations between $[OII] \lambda  \lambda 3727, 3729$, 
$[OIII] \lambda  \lambda 4959, 5007$ line intensities and the oxygen abundance 
have been derived empirically using the available oxygen abundances determined
via measurement of temperature-sensitive line ratios ($T_{e}$ -- method).

In parallel a new $R_{23}$ calibration has been derived on the base of recent 
data and compared to previous calibrations. For oxygen-rich HII regions the 
present $R_{23}$ calibration is close to that of Edmunds and Pagel 
(1984): their calibration has the same slope but is shifted towards higher 
oxygen abundances by around 0.07 dex as compared to the present 
calibration. 
   

\end{abstract}

\section{Introduction}

The oxygen abundance is one of the fundamental characteristics of a galaxy. 
The radial distribution of the oxygen abundance in a galaxy (together with
radial distributions of gas and star surface mass densities) provides a strong
constraint on models of (chemical) evolution. In the general
case the intensities of oxygen emission lines in spectra of HII regions 
depend not only on the oxygen abundance but also on physical conditions.
The oxygen abundance can be derived from accurate measurement of 
temperature-sensitive line ratios, such as   
$[OIII] \lambda \lambda 4959, 5007 / \lambda 4363$. This method 
 will be referred to as the $T_{e}$ -- method. 

Unfortunately, in oxygen--rich HII regions the temperature-sensitive lines like
$[OIII] \lambda 4363$ are too weak to be detected. For such HII regions    
empirical abundance indicators based on more readily observable
lines were suggested (Pagel et al 1979, Alloin et al 1979). The empirical
oxygen abundance indicator $R_{23} = (I_{[OII] \lambda 3727,\lambda 3729}  + 
I_{[OIII] \lambda 4959, \lambda 5007}) /I_{H\beta }$ suggested by Pagel et al
(1979) has found widespread acceptance and use for the oxygen abundance 
determination in HII regions where the temperature-sensitive lines are 
undetectable. This method  will be referred 
to as the $R_{23}$ -- method. Several workers have suggested calibrations of 
$R_{23}$ in terms of oxygen abundance (Edmunds \& Pagel 1984, McCall 
\& Rybski and Shields 1985, Dopita \& Evans 1986, Zaritsky \& Kennicutt and
Huchra 1994, among others). Zaritsky et al's calibration is an average of 
the three former calibrations. 

The oxygen abundances of HII regions in many irregular galaxies are derived
with the $T_{e}$ -- method. Those
 in HII regions of the Milky Way have also been
determined with the $T_{e}$ -- method (Shaver et al 1983). This data
is a base for many investigations of chemical evolution of our Galaxy.
Oxygen abundances in HII regions of many spiral galaxies
have been derived with the $R_{23}$ -- method. As indicated by Skillman et al (1996), 
the precise choice of the O/H -- $R_{23}$ calibration is not critical in 
differential comparisons of the abundance properties of different objects if the 
oxygen abundances in all the objects are derived with the same O/H -- $R_{23}$ 
calibration. However, the comparison of the abundance properties of 
galaxies where the oxygen abundances were determined with the $T_{e}$ -- method
(irregular galaxies, Milky Way  and a few other spiral galaxies) 
and galaxies where the oxygen abundances were determined with the $R_{23}$ -- method
(spiral galaxies) is justified only if the two methods agree.
The existing O/H -- $R_{23}$ 
calibrations (Edmunds \& Pagel 1984, McCall \& Rybski and Shields 1985, 
Dopita \& Evans 1986, McGaugh 1991) are based on then-available oxygen
abundance determinations through the $T_{e}$ -- method and HII region models,
whereas more oxygen abundance determinations through the 
$T_{e}$ -- method are available
now. Then the existing O/H -- $R_{23}$ calibrations should be verified in
the light of more recent data and a new calibration derived if
necessary.

The search for a line intensities -- O/H calibration which results in 
the same oxygen abundances as the $T_{e}$ -- method is the goal of this study.
The line intensities -- O/H calibration has been derived in Section 2. A
discussion will be presented in Section 3. Section 4 is a brief summary.

\section{Line intensities -- O/H calibration}

\begin{figure}[thb]
\vspace{7.5cm}
\includegraphics{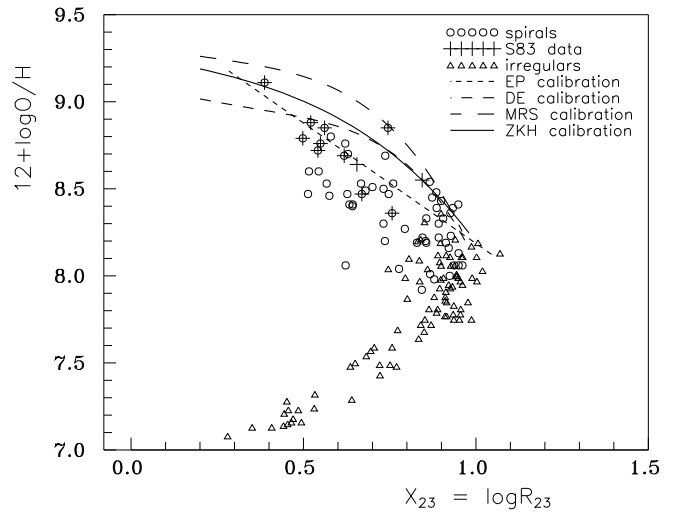}
\caption{\label{figure:9664f1}
The OH -- log$R_{23}$ diagram. The positions of HII regions in spiral galaxies 
(circles) (the HII regions in our Galaxy from Shaver et al 1983 (S83) are 
indicated by plusses) 
and in irregular galaxies (triangles) are shown together with 
OH -- $R_{23}$ calibrations of different authors: Edmunds \& Pagel 1984 (EP),
McCall et al 1985 (MRS), Dopita \& Evans 1986 (DE), and Zaritsky et al 
1994 (ZKH).}
\end{figure}

In order to check whether the  $T_{e}$ -- method and the $R_{23}$ -- method 
result in the same oxygen abundances the $X_{23}$( = log $R_{23}$) versus 
12 + logO/H diagram for the Milky Way Galaxy HII regions from Shaver et al 
(1983) and HII regions in  spiral and irregular galaxies together with 
O/H -- $R_{23}$ calibrations after Edmunds \& Pagel 1984, McCall et al 1985, 
Dopita \& Evans 1986, and Zaritsky et al 1994 has been constructed, 
Fig.\ref{figure:9664f1}. The data for HII regions in irregular galaxies were 
taken from (Izotov and Thuan 1998, 1999: Izotov, Thuan, and Lipovetsky 1994, 
1997; Kobulnicky and Skillman 1996, 
1997, 1998: Kobulnicky et al 1997; Skillman et al 1994;Thuan, Izotov, 
and Lipovetsky 1995; Vilchez and Iglesias-Paramo 1998). 
The data for HII regions in spiral galaxies were taken from 
(Esteban et al 1998;
Esteban et al 1999a,b;
Garnett et al 1999;
Gonzalez-Delgado et al 1995;
Kwitter and Aller 1981;
Pagel, Edmunds, and Smith 1980; 
Peimbert, Torres-Peimbert, and Dufour 1993;
Shaver et al 1983;
Shields and Searle 1978;
van Zee et al 1998;
Vilchez and Esteban 1996;
Vilchez et al 1988;
Webster and Smith 1983).
These lists (involving 151 data points) do not pretend to be
exhaustive ones. In the case of low-metallicity HII regions there is a large 
set of data with recent high-quality determinations of oxygen abundance
with the $T_{e}$ -- method, therefore  earlier ones were not included in
our list. In the case of high-metallicity HII regions 
there are a few recent high-quality determinations of the oxygen abundance
with the $T_{e}$ -- method, therefore we have to include in our list all 
available oxygen abundance determinations although some data were obtained 
around 20 years ago. In a few cases there are two independent determinations of the 
oxygen abundance in the same object or independent determinations 
in different parts of the HII region. Since the goal of the 
present study is a search for the line intensities -- oxygen abundance 
relation but not an investigation of the chemical properties of individual 
galaxies, the independent determinations of oxygen abundance in the same 
object were included in the list as individual data points.
Our data confirm the conclusion of Kennicutt et al (2000) that the 
Edmunds and Pagel calibration (Edmunds \& Pagel 1984) provides a more robust 
diagnostic of oxygen abundance, but an inspection of the 
Fig.\ref{figure:9664f1} shows that it still results in an oxygen abundance 
that is higher than the mean  at any given log$R_{23}$. Thus, none of the 
existing $R_{23}$ -- O/H calibration can reproduce the available data
well enough and a new one should be constructed.

%============================================================Fig f1 (shav-cal)

%============================================================Fig f2 (slope2)
\begin{figure}[thb]
\vspace{7.5cm}
\includegraphics{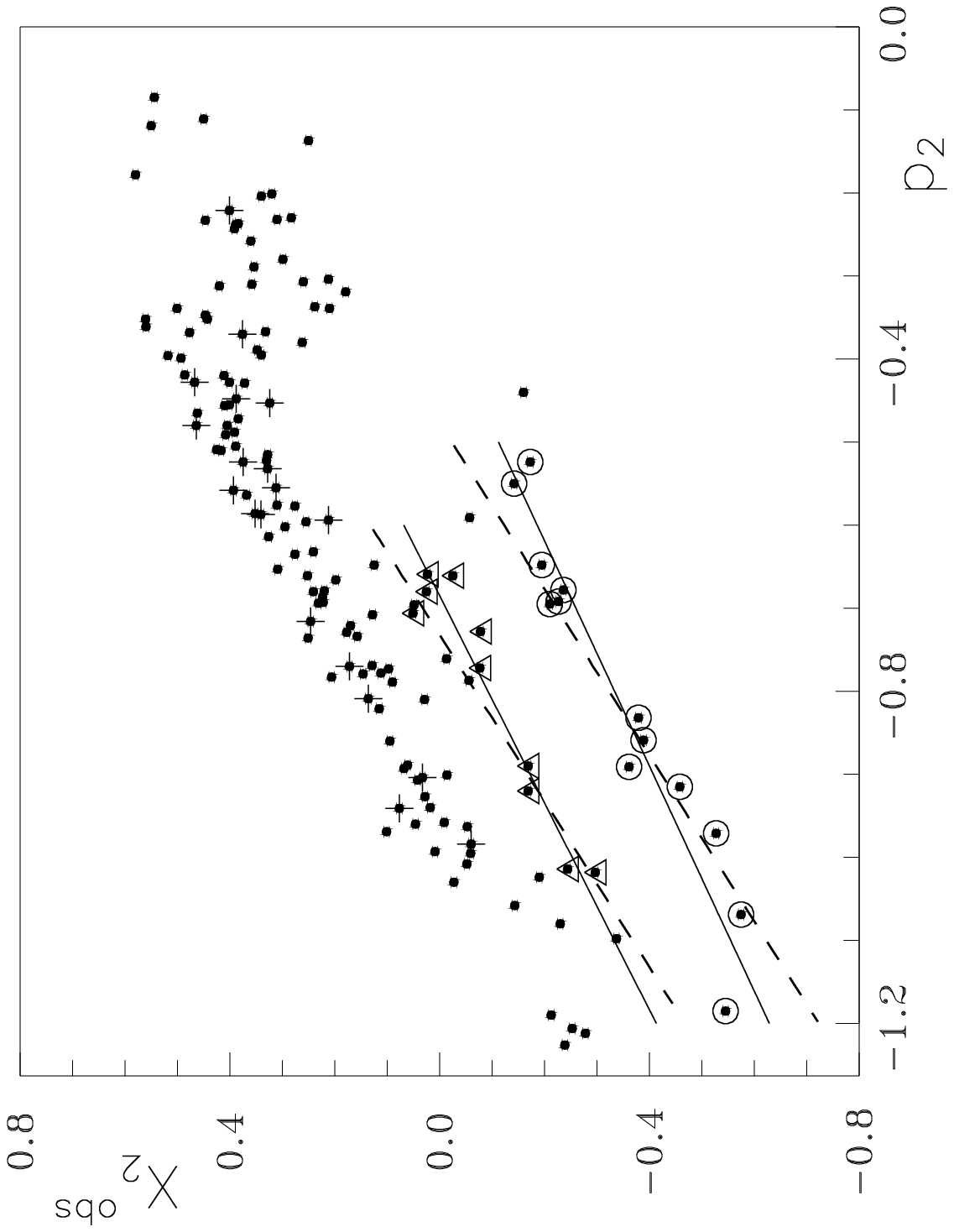}
\caption{\label{figure:9664f2}
The $p_{2}$ versus $X_{2}$ diagram for HII regions in spiral and irregular
galaxies (points). The positions of HII regions with 7.1 $\leq$ 
12+log(O/H)$_{Te}$ $\leq$ 7.3 are shown by circles, HII regions with 7.4 
$\leq$ 12+log(O/H)$_{Te}$ $\leq$ 7.6 are shown by triangles, HII regions with 
8.0 $\leq$ 12+log(O/H)$_{Te}$ $\leq$ 8.1 are shown by plusses. Solid lines
are best fits to positions of HII regions with 7.1 $\leq$ 12+log(O/H)$_{Te}$ 
$\leq$ 7.3 and HII regions with 7.4 $\leq$ 12+log(O/H)$_{Te}$ $\leq$ 7.6.
Dashed lines are lines with slope equal to 1.
}
\end{figure}

%============================================================Fig f3 (slope3)
\begin{figure}[thb]
\vspace{7.5cm}
\includegraphics{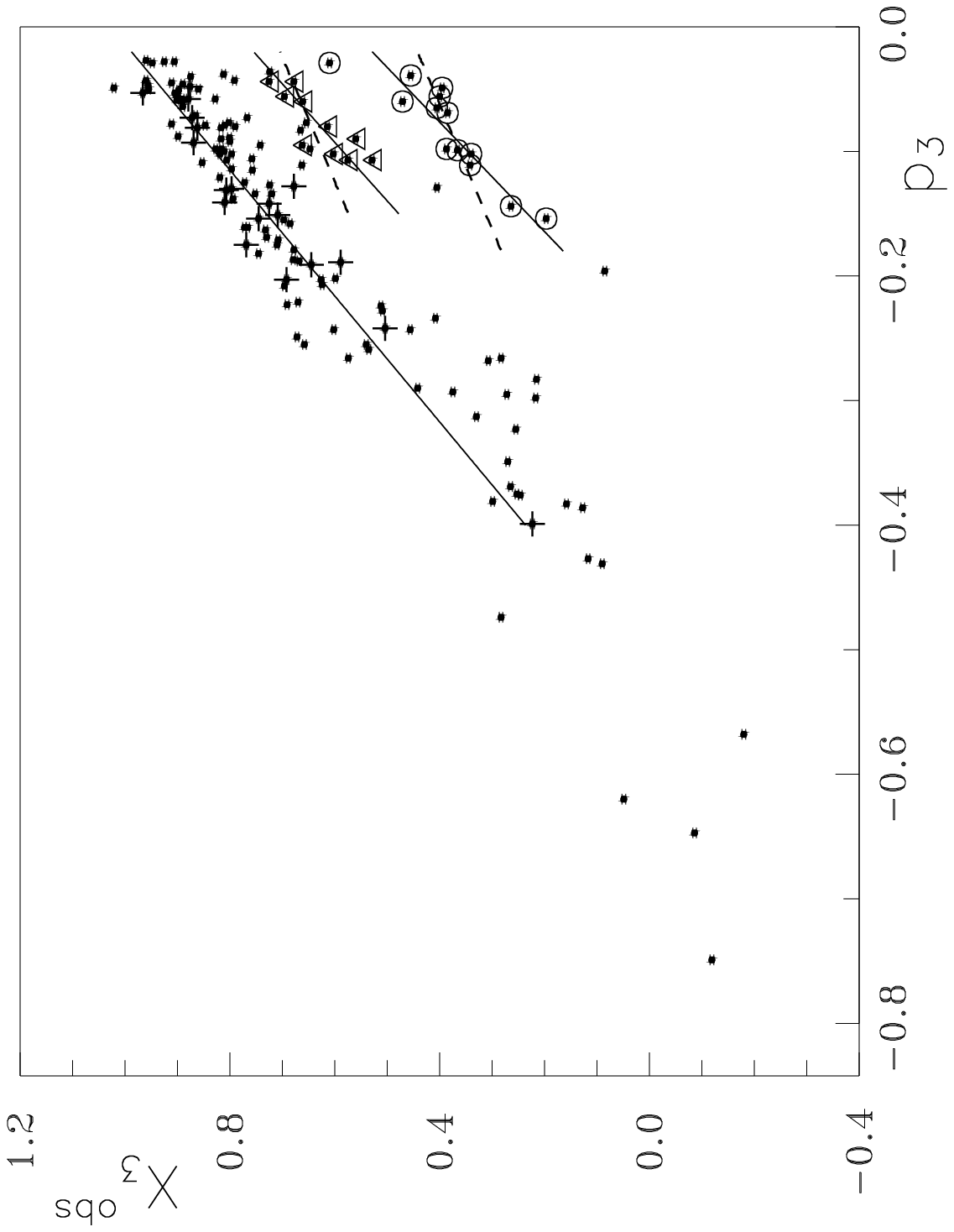}
\caption{\label{figure:9664f3}
The $p_{3}$ versus $X_{3}$ diagram for HII regions in spiral and irregular
galaxies (points). The positions of HII regions with 7.1 $\leq$ 
12+log(O/H)$_{Te}$ $\leq$ 7.3 are shown by circles, HII regions with 7.4 
$\leq$ 12+log(O/H)$_{Te}$ $\leq$ 7.6 are shown by triangles, HII regions with 
8.0 $\leq$ 12+log(O/H)$_{Te}$ $\leq$ 8.1 are shown by plusses. Solid lines
are best fits to positions of HII regions with 7.1 $\leq$ 12+log(O/H)$_{Te}$ 
$\leq$ 7.3, HII regions with 7.4 $\leq$ 12+log(O/H)$_{Te}$ $\leq$ 7.6, and
HII regions with 8.0 $\leq$ 12+log(O/H)$_{Te}$ $\leq$ 8.1.
Dashed lines are lines with slope equal to 1.
}
\end{figure}

%============================================================Fig f4 (xss-OH-i)
\begin{figure}[thb]
\vspace{7.5cm}
\includegraphics{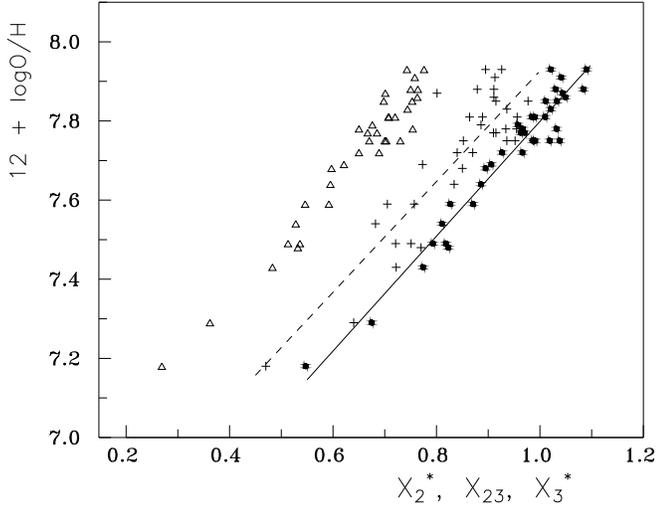}
\caption{\label{figure:9664f4}
The oxygen abundances (O/H)$_{Te}$ versus observed X$_{23}$ and computed
X$_{2}^{*}$ and X$_{3}^{*}$ values for selected subset of galaxies with
best defined oxygen abundances. The O/H versus $X^{*}_{2}$ diagram is 
presented by triangles, the O/H versus $X^{*}_{3}$ diagram is presented by 
points, and the O/H versus $X_{23}$ diagram is presented by plusses
The solid line is the best fit to the O/H versus $X^{*}_{3}$ relation, 
the dashed line is the best fit to the O/H versus $X_{23}$ relation. 
}
\end{figure}

%============================================================Fig f5 (OH-dOH-i)
\begin{figure}[thb]
\vspace{7.5cm}
\includegraphics{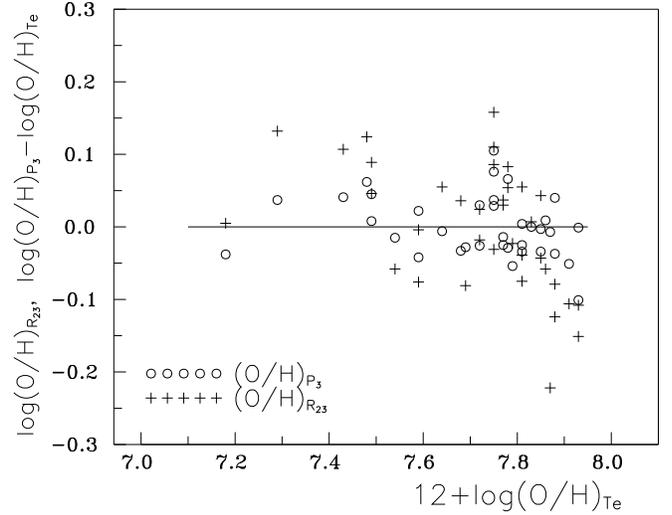}
\caption{\label{figure:9664f5}
The differences between oxygen abundances derived with the suggested p -- method 
and derived with the $T_{e}$ -- method (circles) and
differences between oxygen abundances derived with the $R_{23}$ -- method 
and derived with the $T_{e}$ -- method (crosses) for selected 
subset of HII regions in irregular galaxies with best determined oxygen 
abundances (O/H)$_{Te}$. 
}
\end{figure}

%============================================================Fig f6 (p-dOH-i)
\begin{figure}[thb]
\vspace{7.5cm}
\includegraphics{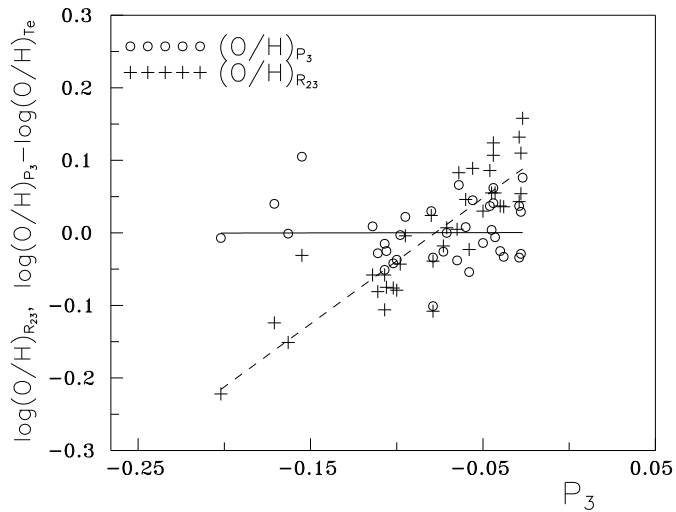}
\caption{\label{figure:9664f6}
The differences between oxygen abundances derived with the suggested p -- method 
and derived with the $T_{e}$ -- method (circles) and
differences between oxygen abundances derived with the $R_{23}$ -- method 
and derived with the $T_{e}$ -- method (plusses) as a 
function of parameter $p_{3}$ for selected subset of HII regions in irregular
galaxies with best determined oxygen abundances through the ${Te}$ -- method.
The solid line is the best fit $\Delta log(O/H)_{P_{3}}$ -- $p_{3}$ relation, 
the dashed line is the best fit $\Delta log(O/H)_{R_{23}}$ -- $p_{3}$ relation. 
}
\end{figure}

%============================================================Fig f7
\begin{figure}[thb]
\vspace{7.5cm}
\includegraphics{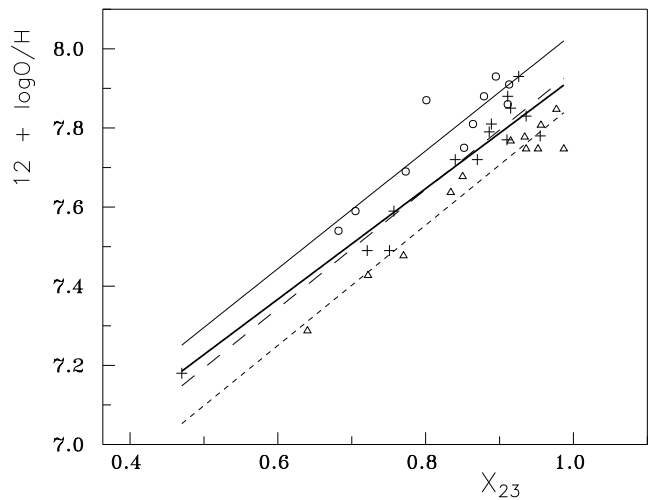}
\caption{\label{figure:9664f7}
The oxygen abundances (O/H)$_{Te}$ versus X$_{23}$ diagram for selected subset 
of galaxies with best defined oxygen abundances. The HII regions with 
$p_{3}$ $<$ --0.1 are shown by circles (and corresponding best fit is presented
by the thin solid line), the HII regions with --0.05 $>$ $p_{3}$ $>$ --0.1 are 
shown by plusses (and corresponding best fit is presented by the long-dashed
line),  the HII regions with $p_{3}$ $>$ --0.05 are shown by triangles (and 
corresponding best fit is presented by the short-dashed line). The thick solid
line is the general best fit, i.e. best fit to all data (the same as in Fig.4). 
}
\end{figure}

%============================================================Fig f8 (xss-OH-t)
\begin{figure}[thb]
\vspace{7.5cm}
\includegraphics{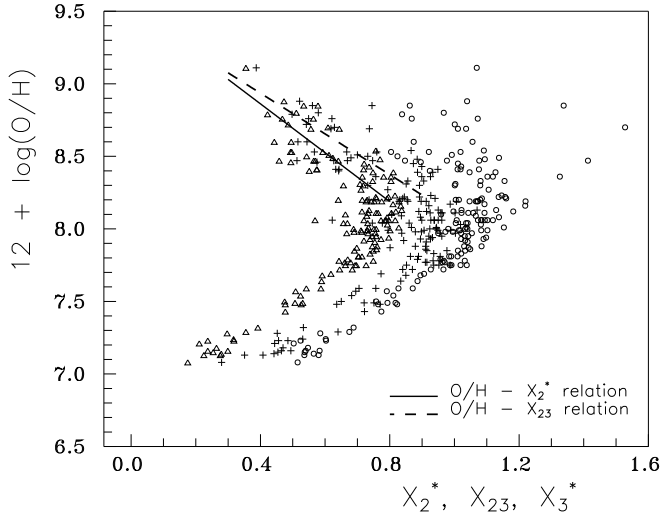}
\caption{\label{figure:9664f8}
The oxygen abundances (O/H)$_{Te}$ versus observed X$_{23}$ (plusses) and 
computed X$_{2}^{*}$ (triangles) and X$_{3}^{*}$ (circles) values for total 
set of HII regions in spiral and irregular galaxies. 
The solid line is the adopted O/H -- $X_{2}^{*}$ 
relation, the dashed line is the adopted O/H -- $X_{23}$ relation.
}
\end{figure}

%============================================================Fig f9(p-dOH-t)
\begin{figure}[thb]
\vspace{7.5cm}
\includegraphics{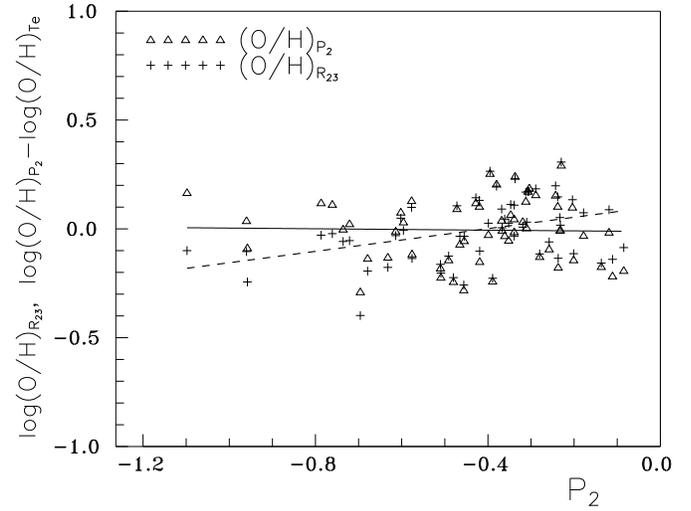}
\caption{\label{figure:9664f9}
The differences between oxygen abundances derived with the suggested p -- method 
and derived with the $T_{e}$ -- method (triangles) and
differences between oxygen abundances derived with the $R_{23}$ -- method 
and derived with the $T_{e}$ -- method (plusses) as a 
function of parameter $p_{2}$.
The solid line is the best fit $\Delta log(O/H)_{P_{2}}$ -- $p_{2}$ relation, 
the dashed line is the best fit $\Delta log(O/H)_{R_{23}}$ -- $p_{2}$ relation. 
}
\end{figure}

%============================================================Fig f10 (x23-OH)
\begin{figure}[thb]
\vspace{7.5cm}
\includegraphics{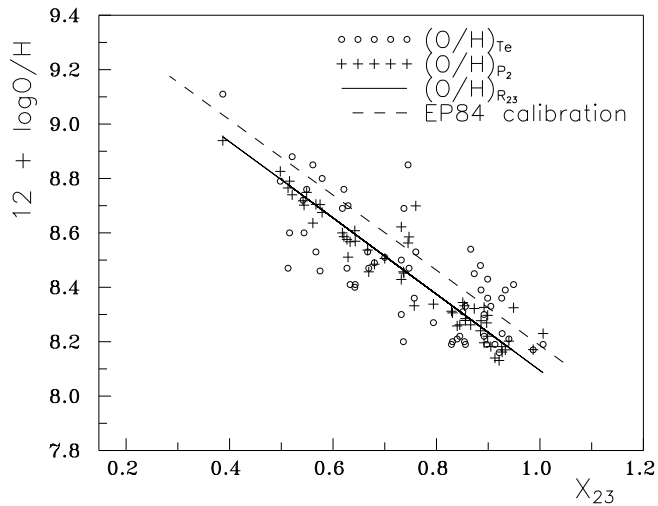}
\caption{\label{figure:9664f10}
The upper branch (12+logO/H $>$ 8.15) of the X$_{23}$ -- O/H diagram. The oxygen 
abundances derived with the $T_{e}$ -- method (original data from literature) are 
shown by circles, the abundances derived with the suggested p -- method are shown 
by plusses. The solid line is the O/H -- $R_{23}$  calibration obtained here,
the dashed line is the O/H -- $R_{23}$  calibration from Edmunds and Pagel (1984).
}
\end{figure}

%============================================================Fig f11 (NGC2403)
\begin{figure}[thb]
\vspace{7.5cm}
\includegraphics{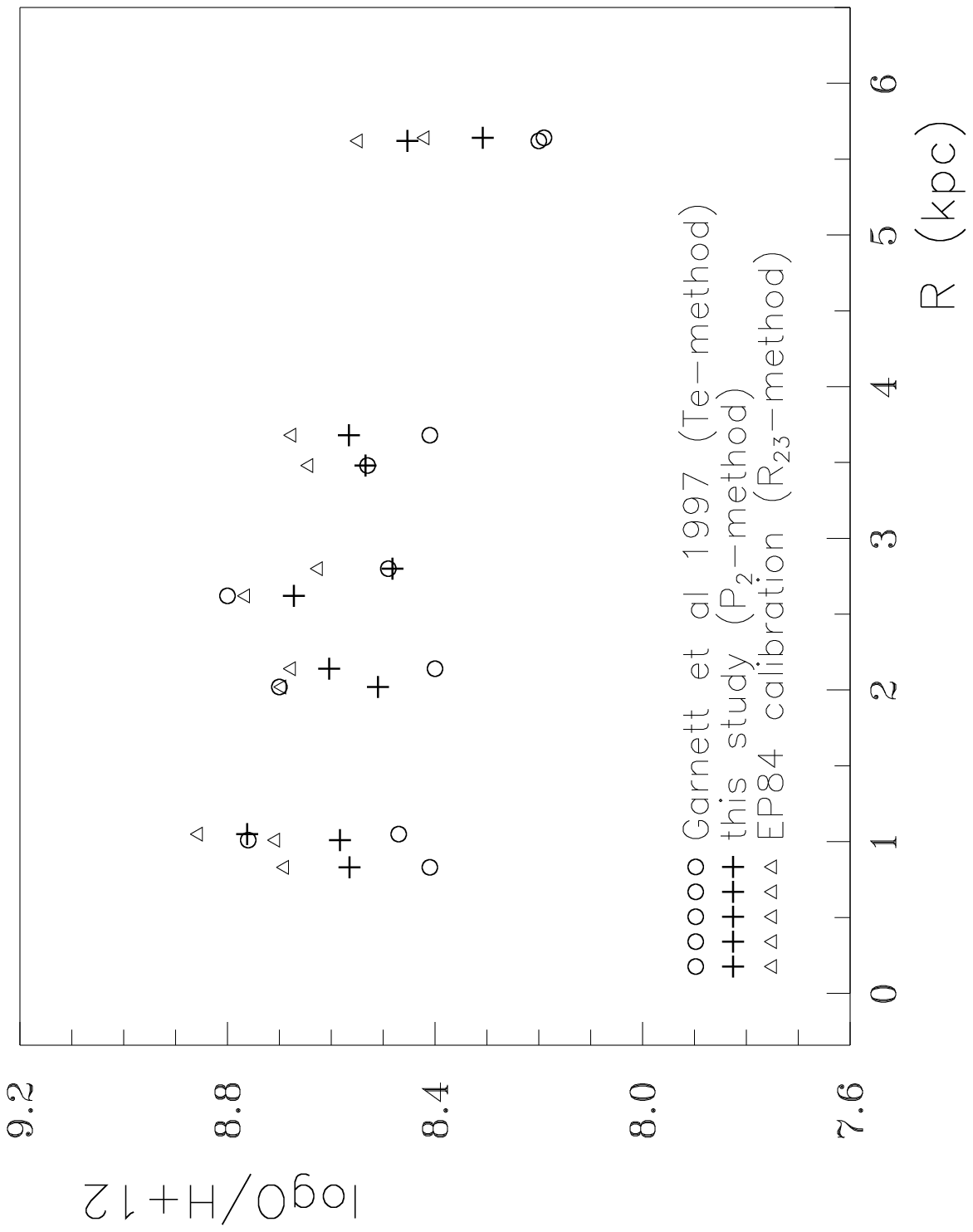}
\caption{\label{figure:9664f11}
The radial distributions of the oxygen abundance in NGC2403 derived by 
Garnett et al (1997) with the $T_{e}$ -- method (circles), derived  
with the p -- method (plusses), and derived with Edmunds and Pagel calibration
($R_{23}$ -- method).
}
\end{figure}

The simplest way to construct the line intensities -- O/H relation is a 
traditional approach: to find a best fit $R_{23}$ -- O/H relation using 
available HII regions in which oxygen abundances were determined with the 
$T_{e}$ -- method. A one-to-one correpondence between the oxygen abundance
and the R$_{23}$ value is implied in this approach, i.e. the variations 
in $R_{23}$ values among HII regions with a given oxygen abundance are not 
taken into consideration. However, McGaugh (1991) pointed out that the 
geometrical factor is important in low-metallicity HII regions and $R_{23}$
must be supplemented with additional information. This can be verified in the 
following simple way.
Let us consider the $X_{2}$ versus $p_{2}$ and $X_{3}$ versus $p_{3}$
diagrams, Figs.\ref{figure:9664f2}, \ref{figure:9664f3}. 
 (The following notations will be accepted here:
$R_{2}$ = $I_{[OII] \lambda 3727+ \lambda 3729} /I_{H\beta }$, 
$X_{2}$ = log$R_{2}$, 
$R_{3}$ = $I_{[OIII] \lambda 4959+ \lambda 5007} /I_{H\beta }$, 
$X_{3}$ = log$R_{3}$, 
$R_{23}$ =$R_{2}$ + $R_{3}$, 
$X_{23}$ = log$R_{23}$, 
$p_{2}$ = $X_{2}$ - $X_{23}$, and
$p_{3}$ = $X_{3}$ - $X_{23}$.)
If the value of $R_{23}$ is constant for HII regions with similar oxygen
abundances, then the HII regions with similar oxygen abundances should
lie along a straight line with a slope equal to 1 in the  $X_{2}$ versus 
$p_{2}$ and $X_{3}$ versus $p_{3}$ diagrams.  The positions of HII
regions with oxygen abundances logO/H+12 in the range from 7.1 to 7.3 are 
shown by circles,  in the range interval from 7.4 to 7.6 are shown by 
triangles, and in the range interval from 8.0 to 8.1 are 
shown by plusses, Figs.\ref{figure:9664f2}, \ref{figure:9664f3}. 
The linear best fits to corresponding data are presented by solid lines. 
Dashed lines are lines with a slope equal to 1. Inspection of 
Figs.\ref{figure:9664f2}, \ref{figure:9664f3} shows that the HII regions 
with similar oxygen abundances lie indeed along a straight line, but
a slope of this straight line is not equal to 1. 

The fact that the slope of the best fit differs from 1 has far-reaching
implications: it means that value of $logR_{23}$ varies systematically with
$p_{3}$ and a quantity other than $logR_{23}$ should be used in 
oxygen abundance determination. In order to verify the reality of this fact 
and to clearly recognize the consequences, the subset of selected
HII regions with best determined oxygen abundances through the $T_{e}$ -- method 
has been considered. This subset includes the low-metallicity 
(12+logO/H $<$ 7.95) HII regions from Izotov \& Thuan (1998, 1999) and 
Izotov \& Thuan, and Lipovetsky (1994, 1997). 
Since the HII regions with similar oxygen abundances lie along a straight line
with a slope other than unity the extrapolated intersect $X_{3}$ is not equal 
to $R_{23}$. The notation X$^{*}_{3}$ will denote the value of $X_{3}$ 
extrapolated to p$_{3}$ = 0. Similarly, the notation X$^{*}_{2}$ will be 
adopted for the value of $X_{2}$ extrapolated to p$_{2}$ = 0.           
The data for the selected subset of HII regions result in the
following relation between $\Delta X_{3}$ = X$_{3}^{obs}$ -- X$^{*}_{3}$ and 
$\Delta$p$_{3}$
\begin{equation}
\Delta X_{3} =  2.20 \; \Delta p_{3} = 2.20 \; p_{3},
\end{equation}
where $\Delta p_{3}$ = p$_{3}$ by virtue of $p^{*}_{3}$ = 0 was adopted.
The corresponding relation between $\Delta X_{2}$ = X$_{2}^{obs}$ -- 
X$^{*}_{2}$ and $\Delta$p$_{2}$ = p$_{2}$ is given by equation
\begin{equation}
\Delta X_{2} =  0.76 \; \Delta p_{2} = 0.76 \; p_{2}.
\end{equation}
Using these equations the values of X$^{*}_{3}$ and X$^{*}_{2}$ have been
computed for all the HII regions from the selected subset. 

The oxygen abundances (O/H)$_{Te}$ versus observed X$_{23}$ and versus
computed X$_{2}^{*}$ and X$_{3}^{*}$ values for the selected subset of HII 
regions are shown in Fig.\ref{figure:9664f4}. 
The O/H versus $X^{*}_{2}$ diagram is 
presented by triangles, the O/H versus $X^{*}_{3}$ diagram is presented by 
points, and the O/H versus $X_{23}$ diagram is presented by plusses
The solid line is the best fit to the O/H versus $X^{*}_{3}$ relation, 
the dashed line is the best fit to the O/H versus $X_{23}$ relation. 
The X$_{23}$ values are 
positioned between corresponding X$^{*}_{2}$ and X$^{*}_{3}$ values. 
Since $p_{3}$ values are more close to zero than $p_{2}$ values  
the extrapolated intersect $X_{3}$ seem to be more reliable than 
 $X_{2}$. Inspection of 
Fig.\ref{figure:9664f4} shows that linear approximations are acceptable
for the relations between O/H and X$_{23}$ and between O/H and X$^{*}_{3}$. 
Thus, the values of X$_{23}$ and X$^{*}_{3}$ have been calibrated in terms of 
oxygen abundance using the linear approximation. The best fits to the 
data result in the following relations (Fig.\ref{figure:9664f4}) 
\begin{equation}
12 + log(O/H)_{R_{23}}  = 6.53 + 1.40 \; X_{23} ,
\end{equation}
\begin{equation}
12 + log(O/H)_{P_{3}}  = 6.35 + 1.45 \; X^{*}_{3} .
\end{equation}
Using these equations two values of the oxygen abundance $(O/H)_{R_{23}}$ and
$(O/H)_{P_{3}}$ have been obtained for every HII region in the selected subset. 
The determination of the oxygen abundance through $p_{3}$ (or $p_{2}$) will be 
referred to as the p -- method.

The differences $\Delta log(O/H)_{R_{23}}$ between oxygen abundances 
$(O/H)_{R_{23}}$ derived with the $R_{23}$ -- method and oxygen abundances 
$(O/H)_{Te}$ derived with the $T_{e}$ -- method are shown in Fig.\ref{figure:9664f5} 
by plusses as a function of $(O/H)_{Te}$. The differences  
$\Delta log(O/H)_{P_{3}}$ = $(O/H)_{P_{3}}$ - $(O/H)_{Te}$  are
are shown in Fig.\ref{figure:9664f5} by circles. 
It can be seen in Fig.\ref{figure:9664f5} that the mean value of 
$\Delta log(O/H)_{P_{3}}$ is appreciable lower ($\sim$ 0.042 dex) than the 
mean value of $\Delta log(O/H)_{R_{23}}$ ($\sim$ 0.086 dex).

The differences $\Delta log(O/H)_{P_{3}}$ and $\Delta log(O/H)_{R_{23}}$ 
as a function of p$_{3}$ are shown in Fig.\ref{figure:9664f6}. As seen in 
Fig.\ref{figure:9664f6} the error in the value of the oxygen abundance 
derived with the $R_{23}$ -- method involves two parts: the first part is a 
random error comparable to the random error in the p -- method, and the 
second part is a systematic 
error. The origin of this systematic error is as follows.
In a general case the intensities of oxygen emission lines in spectra 
of HII regions depend not only on the oxygen abundance but 
also on the physical conditions (hardness of the ionizing 
radiation and geometrical factor). Then in the determination of the
oxygen abundance from line intensities the physical conditions in the HII
region should be taken into account. In the $T_{e}$ -- method this is done
via $T_{e}$. In the p -- method, physical conditions are allowed for  
via parameter p, i.e. the parameter p can be considered as some kind analogy 
of the electron temperature $T_{e}$ in the oxygen abundance determination. In 
the $R_{23}$ -- method the physical conditions in HII region are ignored. 
Therefore, the oxygen abundances derived with the $T_{e}$ -- method and with 
the p -- method involve only random errors (this is a strong argument that 
the physical conditions in HII region are well taken into account 
via parameter p), while the oxygen abundances derived with the $R_{23}$ -- method
involve a systematic error caused by the failure to take into account the
differences in physical conditions in different HII regions. 

The fact that the value of $logR_{23}$ in HII regions with a given oxygen
abundance varies systematically with $p_{3}$ can be directly established 
from detailed consideration of the $X_{23}$ versus O/H diagram, Fig.\ref{figure:9664f7}.
The HII regions with $p_{3}$ $<$ --0.1 are shown by circles (and corresponding 
best fit is presented by the thin solid line) in Fig.\ref{figure:9664f7},
the HII regions with --0.05 $>$ $p_{3}$ $>$ --0.1 are 
shown by plusses (and corresponding best fit is presented by the long-dashed
line),  the HII regions with $p_{3}$ $>$ --0.05 are shown by triangles (and 
corresponding best fit is presented by the short-dashed line). The thick solid
line in Fig.\ref{figure:9664f7} is the general best fit, i.e. the best fit to 
all data (the same as in 
Fig.\ref{figure:9664f4}). Inspection of Fig.\ref{figure:9664f7} shows
that the HII regions with different values of $p_{3}$ lie along different 
straight lines shifted relative to each other depending on the value of $p_{3}$.
According to the model grid of McGaugh (1991) 
this shifting is accounted for by the variations in the geometrical factors.
The general best fit O/H -- X$_{23}$ (equation (3)) is very close to the best 
fit to the data for HII regions with --0.05 $>$ $p_{3}$ $>$ --0.1, 
Fig.\ref{figure:9664f7}.  The HII regions with 
$p_{3}$ $>$ --0.05 are shifted to the right from the general best fit 
and as a consequence the $(O/H)_{R_{23}}$ values derived in
these HII regions with equation (3) are higher than $(O/H)_{Te}$, 
Fig.\ref{figure:9664f6}. Conversely, the HII regions with $p_{3}$ $<$ --0.10 
are shifted to the left from the general best fit in 
Fig.\ref{figure:9664f7}, and so the $(O/H)_{R_{23}}$ derived for them from
equation (3) are lower than $(O/H)_{Te}$, Fig.\ref{figure:9664f6}. 

Thus, the above discussion of the subset of selected HII regions with best 
determined oxygen abundances through the $T_{e}$ -- method suggests that
{\it i)} the oxygen abundances derived with the $R_{23}$ -- method involve a
systematic error caused by the failure to take into account the differences 
in physical conditions in different HII regions, 
{\it ii)} the physical conditions in HII region are well taken into account 
via the parameter p, and there is no systematic error in the oxygen abundances
derived with the p -- method.
We confirm the idea of McGaugh (1991) that the strong oxygen lines
($[OII] \lambda \lambda 3727, 3729$ and $[OIII] \lambda \lambda 4959, 5007$)
contain the necessary information for determination of accurate abundances in
low-metallicity HII regions.

McGaugh (1991) has found that all the models converge toward the same upper
branch line, $R_{23}$ being relatively insensitive to geometrical factor and
ionizing spectra in this region of the diagram.
Using equations (1) and (2) the values of X$^{*}_{3}$ and X$^{*}_{2}$ 
have been computed for all the HII regions from our compilation. The oxygen 
abundances (O/H)$_{Te}$ versus observed X$_{23}$ and versus computed
X$_{2}^{*}$ and X$_{3}^{*}$ values are shown in Fig.\ref{figure:9664f8}. 
The (O/H)$_{Te}$ versus X$_{2}^{*}$ is shown by triangles, the (O/H)$_{Te}$ 
versus X$_{23}$ is presented by plusses, and (O/H)$_{Te}$ versus X$_{3}^{*}$ 
is shown by circles.
The X$_{23}$ values are again positioned between corresponding X$^{*}_{2}$ and 
X$^{*}_{3}$ values. In the case of oxygen-rich HII regions the values of $p_{2}$ 
are usually closer to zero than those of $p_{3}$. Hence, the 
extrapolated intersect $X_{2}$ seems to be more reliable than the extrapolated 
intersect $X_{3}$ and the values of X$^{*}_{2}$ are more suitable for the oxygen 
abundance determination in the oxygen-rich HII regions. The dispersions 
in X$^{*}_{2}$ and X$^{*}_{3}$ values for a given oxygen abundance are 
significantly larger for oxygen-rich than for oxygen-poor HII regions.

For HII regions with 12+log$(O/H)_{Te}$ $>$ 8.15 the relations between O/H and 
X$_{23}$ and  between O/H and X$^{*}_{2}$ have been again  approximated by 
linear relationships. The adopted relation between O/H and X$_{23}$ (dashed 
line in Fig.\ref{figure:9664f8}) is 
\begin{equation}
12 + log(O/H)_{R_{23}}  = 9.50 - 1.40 \; X_{23} ,
\end{equation}
and the adopted relation between O/H and X$^{*}_{2}$ (solid line in 
Fig.\ref{figure:9664f8}) 
\begin{equation}
12 + log(O/H)_{P_{2}}  = 9.54 - 1.68 \; X^{*}_{2} .
\end{equation}

The differences $\Delta log(O/H)_{P_{2}}$ and  $\Delta log(O/H)_{R_{23}}$ 
as a function of p$_{2}$ are shown in 
Fig.\ref{figure:9664f9}.  As in the case of the low-metallicity HII regions,
the error in the value of oxygen abundance derived with the $R_{23}$ -- method 
involves two parts: a random error and 
a systematic error. However in the case of oxygen-rich HII regions the
systematic errors are masked by large random errors. There is no systematic 
error in the oxygen abundances derived with the p -- method.
Thus, the determinations of oxygen abundances with the correction of $R_{23}$ 
for excitation effects result in oxygen abundances which are in better agreement
with oxygen abundances derived through the $T_{e}$ -- method both in low- and 
in high-metallicity HII regions.

\section{Discussion}

Thus, the following simple method of oxygen abundance determination based on 
the readily observable lines $[OII] \lambda \lambda 3727, 3729$ and
$[OIII] \lambda \lambda 4959, 5007$ is suggested. 
Using equation (1) the value of X$_{3}^{*}$ = X$_{3}^{obs}$ -- $\Delta X_{3}$ 
(or using equation (2) the value of X$_{2}^{*}$ = X$_{2}^{obs}$ -- 
$\Delta X_{2}$) is derived, and then using equation (4) for oxygen-poor HII
regions (or equation (6) for oxygen-rich HII regions) the value of 
12 + logO/H is determined. It should be emphasized however that the linear 
approximations of $\Delta X_{2}$ versus p$_{2}$ ($\Delta X_{3}$ versus p$_{3}$)
and O/H versus X$^{*}_{2}$ (O/H versus X$^{*}_{3}$)
relations can be a simplification of reality and more rigorous 
treatment can require the use of more complex curves for these 
relations. It is impossible to find with confidence these curves
with available data and we have to use a linear approximation. Then
the suggested numerical expressions should be considered as a first-order
approximation.

The O/H versus $X_{23}$ diagram for upper branch (12+log(O/H) $>$ 8.15)
is presented in Fig.\ref{figure:9664f10}.  The oxygen abundances 
derived with the $T_{e}$ -- method (original data from literature) are shown by 
circles, those derived with the suggested p -- method are shown by 
crosses, the $R_{23}$ -- O/H calibration derived here is shown by the solid
line, and $R_{23}$ -- O/H calibration after Edmunds and Pagel (1984) is
presented by the dashed line. In the considered range of oxygen abundances the
$R_{23}$ -- O/H calibration of Edmunds and Pagel (1984) can be expressed as
\begin{equation}
12 + log(O/H)_{EP84}  = 9.57 - 1.38 \; X_{23} .
\end{equation}
Inspection of Fig.\ref{figure:9664f10} and comparison of equations (5) and (7) 
shows that the $R_{23}$ -- O/H calibration derived here and that
of Edmunds and Pagel have in fact the same slopes.
The latter is shifted towards 
higher oxygen abundances by about 0.07 dex as compared to the one
 derived here on the basis of more recent data.
Other previous calibrations (McCall \& Rybski and Shields 1985, 
Dopita \& Evans 1986, Zaritsky et al 1994) are shifted towards still
higher oxygen abundances.
 
As can be seen in Fig.\ref{figure:9664f10} there is no one-to-one correspondance 
between $X_{23}$ value and oxygen abundance derived with the p -- method.
Inspection of Fig.\ref{figure:9664f9} shows that the differences between oxygen 
abundances derived with the p -- method and with the $R_{23}$ -- method are
systematically changed with $p_{2}$ from around --0.1 dex for HII regions
with $p_{2}$ $\sim$ --0.1 to around +0.2 dex for HII regions with $p_{2}$ 
$\sim$ --1. In the general case, two HII regions (with $p_{2}$ $\sim$ --0.1 and 
with $p_{2}$ $\sim$ --1) can have the same $(O/H)_{R_{23}}$ while their 
$(O/H)_{P_{2}}$ can differ by $\sim$ 0.3 dex. 
For HII regions with $p_{2}$ from $\sim$ --0.7 to $\sim$ 
--0.2 (majority of HII regions in the present compilation) these differences 
are appreciably less than differences between oxygen abundances derived with 
the $R_{23}$ -- method (or the p -- method) and those derived with the 
$T_{e}$ -- method, Fig.\ref{figure:9664f9}. It is easy to understand that
the proximity of $(O/H)_{R_{23}}$ and $(O/H)_{P_{2}}$ for HII regions with
$p_{2}$ in the range from $\sim$ --0.7 to $\sim$ --0.2 is caused by the fact
that the HII regions with $p_{2}$ values from this range lie close to the 
derived O/H -- X$_{23}$ relation (see discussion for low-metallicity
HII regions and Fig.\ref{figure:9664f7}).

The exactness of oxygen abundance determination with the suggested
p -- method should be estimated. This can be easily done for low - 
metallicity HII regions where there is a large subset of
homogeneous high-quality oxygen abundance determinations with the 
$T_{e}$ -- method. As it was indicated above the average value of
the difference $\Delta log(O/H)_{P_{3}}$
is equal to 0.042 dex (by absolute value) , and maximum value is around
0.1 dex. Thus, the p -- method can be used for oxygen abundance determination
in oxygen-poor HII regions in which the temperature-sensitive lines like 
$[OIII] \lambda 4363$ are measured with large uncertainty or are undetectable.
For oxygen-poor HII regions the exactness of oxygen abundance determination
with the p -- method is comparable with the exactness provided by the 
$T_{e}$ -- method.

The estimation of an exactness of the oxygen abundance determination with the
p -- method in the case of the oxygen-rich HII regions (upper branch) is more 
problematic since few high-quality oxygen abundance determinations with the
$T_{e}$ -- method are available. The mean value of $\Delta log(O/H)_{P}$ 
(by absolute value) for oxygen-rich HII regions is 0.14, and for an individual 
HII region this difference can be as large as around 0.25, 
Fig.\ref{figure:9664f9}. In the limiting case, when it is assumed that the 
oxygen abundances derived with the $T_{e}$ -- method are precise, the values of 
$\Delta log(O/H)_{P}$ are totally accounted for by the inexactness of the
p -- method. This seems not to be the case, as it is well known that the 
exactness of abundance 
determinations in metal-rich HII regions is rather low (see present-day review 
in Henry and Worthey 1999). The temperature-sensitive lines like 
$[OIII] \lambda 4363$ used in the $T_{e}$ -- method are very weak in oxygen-rich 
HII regions and are measured with large uncertainty that results in large 
uncertainty in the oxygen abundances. Then the uncertainties in the oxygen 
abundances derived with the $T_{e}$ -- method can make a significant (may be a
dominant) contribution 
to $\Delta log(O/H)_{P}$. In this case  it can be suggested (without pretending 
that the p-method provides more accurate oxygen abundances than the
$T_{e}$ -- method in principle) that the p -- method (and the $R_{23}$ -- method for
HII regions with low level of excitation) provides as accurate oxygen 
abundances in oxygen-rich HII regions as
the $T_{e}$ -- method taking into account the present-day state-of-art with the
line intensity measurements. It should be noted that if errors in $(O/H)_{Te}$
are random they impact weakly on the $X^{*}_{2}$ -- O/H and $X_{23}$ -- O/H 
calibrations based on the $(O/H)_{Te}$.

The biggest set of HII regions in individual galaxy with oxygen abundances 
determined with the $T_{e}$ -- method is the data of Garnett et al (1997) for HII 
regions in NGC2403. Fig.\ref{figure:9664f11} shows the radial distributions 
of oxygen abundance in NGC2403 derived by Garnett et al (1997) with the $T_{e}$ -- 
method (circles), derived with the p -- method (plusses), and derived with the 
$R_{23}$ -- method (Edmunds and Pagel calibration) (triangles).
Inspection of Fig.\ref{figure:9664f11} shows that the oxygen abundances 
derived with the p -- method correlates even more tightly with the 
galactocentric distance than the oxygen abundances derived with the $T_{e}$ -- 
method. Since the level of excitation in HII regions of NGC2403 is not very
high ($p_{3}$ is less than -0.2 for any HII region) the $R_{23}$ -- method
(present calibration) results in the oxygen abundances which are close to
that derived with the p -- method. 
The oxygen abundances determined with calibration of Edmunds and 
Pagel are slightly higher than that obtained with present calibration, 
in agreement with above
conclusion that the $R_{23}$ -- O/H calibration of Edmunds and Pagel is shifted 
towards higher oxygen abundances by around 0.07 dex. 
Thus, the case of NGC2403 confirms that the p -- method and the $R_{23}$ -- method
provide as accurate oxygen abundances in oxygen-rich HII regions as
the $T_{e}$ -- method.

\section{Conclusions}

The problem of line intensities -- oxygen abundance calibration has been
considered. It has been obtained that
the oxygen abundances derived with the $R_{23}$ -- method involve a
systematic error caused by the failure to take into account the differences 
in physical conditions in different HII regions.

We confirm the idea of McGaugh (1991) that the strong (readily observable) 
oxygen lines ($[OII] \lambda \lambda 3727, 3729$ and 
$[OIII] \lambda \lambda 4959, 5007$)
contain the necessary information for determination of accurate abundances in
low-metallicity (and may be also in high-metallicity) HII regions.
It has been found that the excitation parameters $p_{3}$ 
or $p_{2}$ (which are defined here as contributions of the radiation in 
$[OIII] \lambda \lambda 4959, 5007$ lines and in 
$[OII] \lambda \lambda 3727, 3729$ lines to the "total" oxygen radiation 
respectively) allow to take into account the variations in $R_{23}$ values
among HII regions with a given oxygen abundance.
Based on this fact a new way of the oxygen abundance determination in
HII regions (p -- method) has been constructed and corresponding 
relations between $[OII] \lambda  \lambda 3727, 3729$, 
$[OIII] \lambda  \lambda 4959, 5007$ line intensities and the oxygen abundance 
have been derived empirically using the available oxygen abundances determined
via measurement of temperature-sensitive line ratios ($T_{e}$ -- method).

By comparing of oxygen abundances in HII regions derived with the
$T_{e}$ -- method and those derived with the p -- method
it has been found that for the low-metallicity HII regions the exactness 
of oxygen abundance determination with the p -- method is comparable to that
obtained with the $T_{e}$ -- method.
For the low-metallicity HII regions the p -- method provides a more robust 
diagnostic of oxygen abundance than the $R_{23}$ -- method.

For oxygen-rich HII regions both p and $R_{23}$ calibrations are less reliable.
For the majority of HII regions in the present compilation  the differences 
between oxygen abundances derived with the p -- method and with the 
$R_{23}$ -- method are appreciably less than between those derived with the 
$R_{23}$ -- method (or the p -- method) and those from the 
$T_{e}$ -- method. In the general case, however, two HII regions can have the 
same $(O/H)_{R_{23}}$ while their $(O/H)_{P_{2}}$ can differ by $\sim$ 0.3 dex. 

The calibration of $R_{23}$ presented here is compared to previous calibrations.
The calibration of Edmunds and Pagel (1984) has the same slope but
is shifted towards higher oxygen abundances by about 0.07 dex as 
compared to the present calibration. Other previous calibrations (McCall \& 
Rybski and Shields 1985, Dopita \& Evans 1986, Zaritsky et al 1994) are shifted 
towards still higher abundances.

For the high-metallicity HII regions the exactness of oxygen abundance 
determination with the suggested p -- method cannot be firmly estimated 
due to the lack of the high-quality determinations of oxygen abundances 
with the $T_{e}$ -- method. Indirect arguments suggest that the p -- method 
provides as accurate oxygen abundances in oxygen-rich HII regions as
the $T_{e}$ -- method taking into account the present-day state-of-art with 
the line intensity measurements.

\begin{acknowledgements}
I thank the referee, Prof. B.E.J.Pagel, for helpful comments and suggestions
as well as improving the English text.
This study was partly supported by the INTAS grant No 97 -- 0033 and the
NATO grant PST.CLG.976036. 
\end{acknowledgements}

\end{document}